\newcommand*{\mathspace}{~~~~~}

\documentclass[%
 jmp,
 amsmath,amssymb,
preprint,%
nofootinbib
]{revtex4-1}

\usepackage{graphicx}% Include figure files
\usepackage{dcolumn}% Align table columns on decimal point
\usepackage{bm}% bold math
\usepackage{hyperref}
\usepackage{bbold}

\usepackage[utf8]{inputenc}
\usepackage[T1]{fontenc}
\usepackage{mathptmx}
\usepackage{braket}
\usepackage{siunitx}
\usepackage{color}
\usepackage{comment}

\begin{document}

%\preprint{???}

\title[Two particles interacting on $S^2$]{Two particles interacting via a contact interaction on $S^2$}
% Force line breaks with \\

\author{D. Schuh}
\email{schuh@hiskp.uni-bonn.de}
 \affiliation{Helmholtz-Institut f\"ur Strahlen- und Kernphysik, Rheinische Friedrich-Wilhelms-Universit\"at Bonn, 53012 Bonn, Germany}%Lines break automatically or can be forced with \\
\author{T. Luu}%
 \email{t.luu@fz-juelich.de}
\affiliation{ 
Institut f\"ur Kernphysik and Institute for Advanced Simulation, Forschungszentrum J\"ulich, 54245 J\"ulich, Germany
}%
%\affiliation{Helmholtz-Institut f\"ur Strahlen- und Kernphysik, Rheinische Friedrich-Wilhelms-Universit\"at Bonn, 53012 Bonn, Germany}

%\author{C. Author}
% \homepage{http://www.Second.institution.edu/~Charlie.Author.}
%\affiliation{%
%Second institution and/or address%\\This line break forced% with \\
%}%

\date{\today}% It is always \today, today,
             %  but any date may be explicitly specified

\begin{abstract}
We consider two particles interacting via a contact interaction that are constrained to a sphere, or $S^2$.  We determine their spectrum to arbitrary precision and for arbitrary angular momentum.    We show how the non-inertial frame leads to non-trivial solutions for different angular momenta.  Our results represent an extension of the finite-volume L\"uscher formulas but now to a non-trivial geometry.  We apply our results to predict the spectrum of select two-nucleon halo nuclei and compare with experimental results.
\end{abstract}

\maketitle

\section{\label{sec:intro}Introduction}
The eigenvalue solutions to two interacting particles is a standard topic introduced to beginning students of quantum mechanics.  Typical first examples include two particles interacting via a contact interaction and the Coulombic solutions of oppositely charged particles.  These examples serve as a stepping stone to more complicated quantum mechanical \emph{many-body} systems whose solutions are usually \emph{not} known.

Besides serving a great pedagogical introduction to many-body quantum mechanics, the two-body system itself plays an important role in multiple fields of physics.  For example, when the particles are placed within a finite cubic volume their eigenvalue solutions satisfy L\"uscher's quantization formula \cite{Luscher:1986pf,Luscher:1990ux,Beane:2003da,Luu:2011ep}.  Lattice Quantum Chromodynamics (LQCD) calculations of composite two-body systems within a finite volume utilize this relation to extract \emph{infinite-volume} interaction parameters between these particles \cite{Briceno:2013lba,Briceno:2013bda}.  Sometimes the finite-volume is dictated by the experimental setup as opposed to numerical convenience, as is the case with cold-ion traps.  Here the confinement of the two particles can be satisfactorily approximated by an external harmonic oscillator well.  Again, the energy solutions here \cite{Busch:1998,Luu:2010hw} provide information on the interacting properties of the particles within this confinement, and in particular whether the two particles undergo a Feshbach resonance when their scattering length diverges \cite{Kohler:2006zz}.  As a final example, solutions exist for two interacting particles within a hard spherical wall \cite{Lu:2015riz}, providing a means for ``tuning" interaction parameters for many-nucleon simulations using nuclear lattice effective field theory \cite{Alarcon:2017zcv, B1_Lahde:2019npb}. 

The examples above refer to systems residing in three spatial dimensions.  But in all these cases there are corresponding solutions in both one- and two-dimensions.  A tacit assumption here is that the interaction between the two particles only depends on their relative coordinates.  When this is the case, and if the geometry allows it, one can readily separate the system into its relative (Jacobi) and center-of-mass (CM) coordinates.  This provides a great simplification to the eigenvalue solutions since one can work solely within the inertial frame.  

In this paper we consider two particles \emph{confined} to a sphere of arbitrary radius $R$ (i.e. confined to $S^2$) and that interact via a contact interaction.  Though this interaction again only depends on the relative coordinates, the surface $S^2$ is a \emph{non-inertial} frame that affords no general separation of relative and CM coordinates and as such there is no simplification to the eigenvalue solution.\footnote{The exception is the case with zero total angular momentum~\cite{doi:10.1063/1.449011}.} Yet we show how solutions of arbitrary precision to this system can be found.  Furthermore, because of the non-inertial frame, we find an infinite tower of solutions depending on the \emph{total} angular momentum $L$.  Our results are not completely academic--we consider two-nucleon halo nuclei and show how our results can be used to extract interaction parameters between the loosely bound nucleons.  We view this system as another excellent pedagogical example of two quantum mechanical interacting particles, but this time within a non-inertial frame.  

Our paper is organized as follows.  In Sect.~\ref{sec:definition} we define our problem.  We derive the quantization conditions for arbitrary total angular momentum $L$ in terms of direct summations in Sect.~\ref{sec:Zfunctions}.  When considering explicit values for $L$, we find analytic expressions for the summations, which we provide for $L=0$ up to $L=2$ in Sect.~\ref{sec:examples}.  We stress, however, that all solutions for $L>2$ can be readily found using our method.  We then apply our results to extract interaction parameters of two-nucleon halo nuclei in Sect.~\ref{sec:halo nuclei}.  We recapitulate in Sect.~\ref{sec:conclusion}.  Detailed and lengthy derivations are reserved for the appendices.

\section{\label{sec:definition}Problem setup}
\begin{figure}
    \centering
    \includegraphics[width=.5\columnwidth]{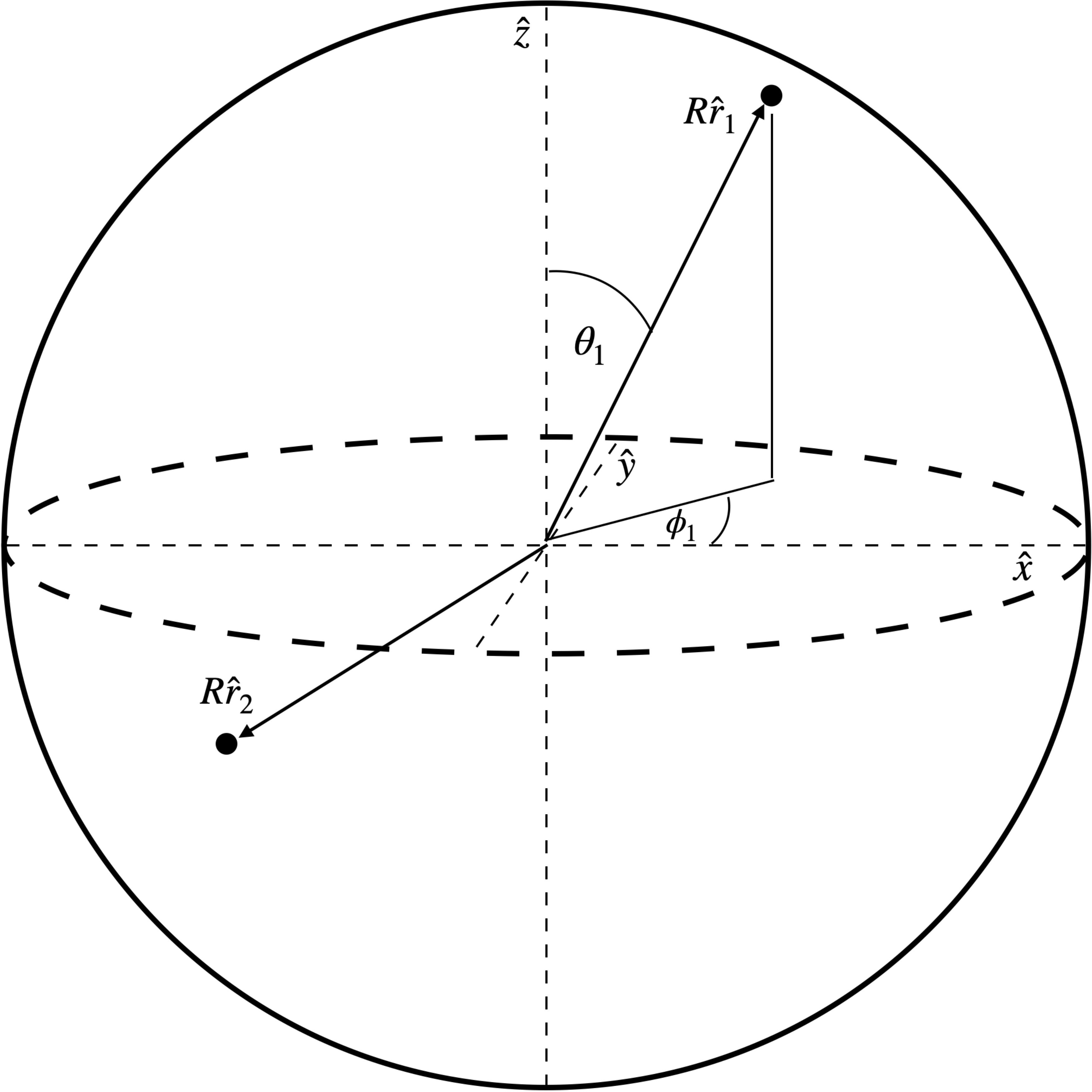}
    \caption{Two particles confined to the surface of a sphere of radius $R$. The internal angular coordinates are only shown for the first particle to reduce clutter.}
    \label{fig:setup}
\end{figure}
We consider two particles of equal mass $m$ confined to the surface of ball of radius $R$, as shown in Fig.~\ref{fig:setup}.  The particles' positions are then solely dictated by their angles $\hat r_1$ and $\hat r_2$, which in turn can be expanded in a basis of spherical harmonics $\langle \hat r|l,m_l\rangle = Y_{lm_l}(\hat r)$.  The kinetic term of the Hamiltonian describing such particle movement is well known and is that of a rigid motor,\footnote{We set $\hbar=c=1$ in all our expressions.}
\begin{equation}\label{eqn:rigid motor hamiltonian}
    \hat T|l,m_l\rangle = \frac{l(l+1)}{2m R^2}|l,m_l\rangle\equiv \epsilon_l|l,m_l\rangle\ .
\end{equation}

\subsection{The contact interaction}
We assume that the particles interact via a contact interaction only, which in this geometry is given in coordinate space by
\begin{equation}\label{eqn:interaction}
\hat V_{12}|\hat {\bm r}_1;\hat {\bm r}_2\rangle=|\hat {\bm r}_1;\hat {\bm r}_2\rangle\frac{C_0(\Lambda)}{R^2}\delta\left(\hat {\bm r}_1-\hat{\bm r}_2\right)
=|\hat {\bm r}_1;\hat {\bm r}_2\rangle\frac{C_0(\Lambda)}{R^2}\delta(\cos\theta_1-\cos\theta_2)\delta(\phi_1-\phi_2)\ .
\end{equation}
Here $C_0(\Lambda)$ is a coefficient that is tuned to reproduce a particular observable of the two-particle system and the variable $\Lambda$ represents a momentum cutoff scale.  The procedure for tuning this coefficient is non-trivial but has been done previously in \cite{Beane:2010ny,Korber:2019cuq}, and we only mention some salient features of this procedure relevant to our analysis in Sect.~\ref{sec:halo nuclei}.   For a more thorough description of this tuning we recommend the reader consult the aforementioned references.

The relevant physical observable is the $s$-wave scattering length $\tilde a$, which in two dimensions is \emph{dimensionless}, despite its name \cite{Hammer:2010fw}.  To a certain degree, the magnitude and sign of this parameter dictates how strongly the particles repulsively or attractively interact with one other. We can define a `reduced scattering length' $a$ that is \emph{dimensionful} by introducing an arbitrary length scale.  We set this length scale to be the radius $R$ of our sphere.  The relation between the physical scattering length $\tilde a$ (dimensionless) and reduced scattering length $a$ (dimension of length) is given by \cite{Hammer:2010fw}
\begin{equation}\label{eqn:reduced scattering length}
    a=R\ \exp\left(-\frac{\pi}{2 \tilde a}\right)\ .
\end{equation}
Note that this definition implies that $a\ge0$.  The tuning of $C_0(\Lambda)$ then follows the procedures described in \cite{Beane:2010ny,Korber:2019cuq}.  Assuming a hard-cutoff regulator in momentum space, the coefficient is
\begin{equation}\label{eqn:C0}
    C_0(\Lambda)=-\frac{2 \pi}{m \log \left(a \Lambda\right)}\ .
\end{equation}
We note that though the interaction Eq.~\eqref{eqn:interaction} is both cutoff and scheme dependent by virtue of the coefficient in Eq.~\eqref{eqn:C0}, observables are not.  We ultimately take the limit $\Lambda\to\infty$ in all our subsequent calculations.

\subsection{The integral equation}
Our task then is to solve the eigenvalue equation\footnote{The eigenvalue $E$ includes both rotational and vibrational energies.}
\begin{equation}\label{eqn:schroedinger}
    \left(\hat{T}_{1}+\hat{T}_{2}+\hat{V}_{12}\right)|\psi_{LM}\rangle=E|\psi_{LM}\rangle\ ,
\end{equation}
where our eigenstates are states with good total angular momentum $L$ and $M$ since our interaction preserves total angular momentum.
To do this, we first recast Eq.~\eqref{eqn:schroedinger} into integral form,
\begin{equation}
    |\psi_{LM}\rangle = \frac{1}{E-\hat{T}_1-\hat{T}_2}\hat{V}_{12}|\psi_{LM}\rangle\ .
\end{equation}
We then project the eigenstate onto
\begin{displaymath}
\bra{(l_1l_2)LM}\equiv\sum_{m_1,m_2}\langle l_1,m_1;l_2,m_2|LM\rangle \langle l_1,m_1|\langle l_2,m_2|\ ,
\end{displaymath}  
where $\langle l_1,m_1;l_2,m_2|LM\rangle$ is a Clebsch-Gordan coefficient.  This gives
\begin{align}\label{eqn:schroedinger_proLected}
    \braket{(l_1l_2)LM|\psi_{LM}} = %\lim \limits_{\Lambda \rightarrow \infty} 
    \frac{1}{E-\epsilon_{l_1}-\epsilon_{l_2}}\sum_{l'_1 l'_2} \Braket{(l_1l_2)LM|\hat{V}_{12}|(l_1'l_2')LM}\Braket{(l_1'l_2')LM|\psi_{LM}} \ .
\end{align}
On the RHS above we have inserted the closure relation $\hat{1}=\sum_{l'_1l'_2}|(l_1'l_2')LM\rangle\langle(l_1'l_2')LM|$ and used the fact that $|l_1m_1\rangle$ and $|l_2m_2\rangle$ are eigenstates of $\hat{T}_1$ and $\hat{T}_2$, respectively, with eigenenergies given in Eq.~\eqref{eqn:rigid motor hamiltonian}.

\subsection{\label{sec:Zfunctions}Quantization condition for general $L$}
To continue further we require the explicit form of the matrix element $\Braket{(l_1l_2)LM|\hat{V}_{12}|(l_1'l_2')LM}$. As this derivation is quite tedious, we leave it for the appendices (App.~\ref{sec:matrix element}) and only provide the end result here:
\begin{multline}\label{eq:matrix_element}
	\Braket{(l_1l_2)LM | V_{12} | (l'_1l'_2)LM} =\\ 
	\begin{cases}
	\displaystyle\frac{C_0(\Lambda)}{R^2}\frac{\sqrt{\hat{l_1} \hat{l_2} \hat{l'_1} \hat{l'_2}}}{4\pi} 
	 \begin{pmatrix}
		l_1 & l_2 & L\\
		0&0&0
	\end{pmatrix}
	\begin{pmatrix}
	l'_1 & l'_2 & L \\
	0&0&0
	\end{pmatrix}\mathspace \forall\mathspace l_i(l_i+1),\ l'_i(l'_i+1)\le(\Lambda R)^2\\
	0 \mathspace\text{otherwise}
	\end{cases}\ ,
\end{multline}
where
\begin{displaymath}
\begin{pmatrix}
		l_1 & l_2 & L\\
		m_1&m_2&M_L
	\end{pmatrix}
	\end{displaymath}
is a Wigner 3-$j$ symbol \cite{edmonds1996angular} and we define $\hat{l_i}\equiv 2l_i+1$ for brevity.  The condition that $l_i(l_i+1)\le(\Lambda R)^2$ and $l'_i(l'_i+1)\le(\Lambda R)^2$ for $i=1,2$ comes from the momentum hard cutoff condition of our interaction.
We now plug this expression into Eq.~\eqref{eqn:schroedinger_proLected}, giving
\begin{multline}\label{eqn:schroedinger_proLected_v2}
    \braket{(l_1l_2)LM|\psi_{LM}} = %\lim \limits_{\Lambda \rightarrow \infty} 
    \frac{C_0(\Lambda)}{R^2}\frac{\sqrt{\hat{l_1} \hat{l_2}}}{4\pi}\begin{pmatrix}
		l_1 & l_2 & L\\
		0&0&0
	\end{pmatrix}\frac{1}{E-\epsilon_{l_1}-\epsilon_{l_2}}\\
	\sum_{l'_1 l'_2}^{\Lambda R}
	\sqrt{ \hat{l'_1} \hat{l'_2}}
	\begin{pmatrix}
	l'_1 & l'_2 & L \\
	0&0&0
	\end{pmatrix}
	  \Braket{(l_1'l_2')LM|\psi_{LM}}\ .
\end{multline}
The equality above holds for all $l_i$ such that $l_i(l_i+1)\le(\Lambda R)^2$ for $i=1,2$.  In particular, it holds if we multiply both sides of the equation by $\sqrt{\hat{l_1}\hat{l_2}}\begin{pmatrix}
		l_1 & l_2 & L\\
		0&0&0
	\end{pmatrix}$ and then \emph{sum both sides over $l_1$ and $l_2$},
\begin{multline}\label{eqn:schroedinger_proLected_v3}
  \sum_{l''_1,l''_2}^{\Lambda R} \sqrt{ \hat{l''_1} \hat{l''_2}}
	\begin{pmatrix}
	l''_1 & l''_2 & L \\
	0&0&0
	\end{pmatrix} \braket{(l''_1l''_2)LM|\psi_{LM}} = %\lim \limits_{\Lambda \rightarrow \infty} 
     \frac{C_0(\Lambda)}{4\pi R^2}\sum_{l_1,l_2}^{\Lambda R}\frac{\hat{l_1} \hat{l_2}}{E-\epsilon_{l_1}-\epsilon_{l_2}}\begin{pmatrix}
		l_1 & l_2 & L\\
		0&0&0
	\end{pmatrix}^2
	\\
	\sum_{l'_1 l'_2}^{\Lambda R}
	\sqrt{ \hat{l'_1} \hat{l'_2}}
	\begin{pmatrix}
	l'_1 & l'_2 & L \\
	0&0&0
	\end{pmatrix}
	  \Braket{(l_1'l_2')LM|\psi_{LM}}\ .
\end{multline}
On the LHS above we have introduced new summation indices $l''_i$ (instead of $l_i$) to stress that it is the sum that holds under the equality.  A trivial solution to the equality occurs if $\braket{(l_1l_2)LM|\psi_{LM}}=0$ for all $l_i$.  To obtain a non-trivial solution, we collect the components $\braket{(l_1l_2)LM|\psi_{LM}}$ to one side of the equation,
\begin{multline}
    0=\\
    \sum_{l'_1 l'_2}^{\Lambda R}
	\sqrt{ \hat{l'_1} \hat{l'_2}}
	\begin{pmatrix}
	l'_1 & l'_2 & L \\
	0&0&0
	\end{pmatrix}
	  \Braket{(l_1'l_2')LM|\psi_{LM}}
	  \left[\frac{C_0(\Lambda)}{4\pi R^2}\sum_{l_1,l_2}^{\Lambda R}\frac{\hat{l_1} \hat{l_2}}{E-\epsilon_{l_1}-\epsilon_{l_2}}\begin{pmatrix}
		l_1 & l_2 & L\\
		0&0&0
	\end{pmatrix}^2-1 \right]\ .
\end{multline}
The equality now holds non-trivially if the term in square brackets vanishes.  Using the exact form of $C_0(\Lambda)$ from Eq.~\eqref{eqn:C0} and equating the term in square brackets to zero gives the desired quantization condition for arbitrary total angular momentum $L$:
\begin{equation}\label{eqn:tada}
\boxed{
\log\left(\frac{a}{R}\right)=%\lim_{\Lambda\to\infty}
\sum_{l_1,l_2}^{\Lambda R}\frac{(2l_1+1)(2l_2+1)}{l_1(l_1+1)+l_2(l_2+1)-x}\begin{pmatrix}
l_1 & l_2 &L\\ 0 & 0 & 0
\end{pmatrix}^2-\log(\Lambda R)}\ ,
\end{equation}
where $x\equiv 2m ER^2$. The summation above is over all $l_i$ such that $l_i(l_i+1)\le(\Lambda R)^2$. %, and we have explicitly put in the limit $\Lambda\to\infty$ on the RHS.

The eigenvalues $E$, or equivalently $x$, of Eq.~\eqref{eqn:schroedinger} are those that satisfy the equality in Eq.~\eqref{eqn:tada}. This represents L\"uscher's formula on $S^2$ for each rotational band $L$ under the assumption of a pure contact interaction.

\subsection{\label{sec:examples}Closed expressions for select $L$}
When we concentrate on specific values of $L$ and take the limit $\Lambda\to\infty$ we can further simplify Eq.~\eqref{eqn:tada} and obtain closed expressions.  We do this explicitly for $L=0$ and $1$, and provide the closed expression for $L=2$ without derivation.  In principle it is possible to obtain closed expressions for concrete values of $L>2$, but the derivation becomes much more tedious and onerous.

\subsubsection{$L=0$}
To start, note that the sums over $l_1$ and $l_2$ in Eq.~\eqref{eqn:tada} are restricted by the triangle inequalities of the Wigner 3-$j$ symbol
\begin{align}
    |l_1 - l_2| \leq L \leq l_1 + l_2\ .
\end{align}
For $L=0$ this implies that $l_1=l_2\equiv l$.  The 3-$j$ symbol simplifies to $\frac{(-1)^l}{\sqrt{2 l+1}}$ and Eq.~\eqref{eqn:tada} becomes
\begin{equation}
\log\left(\frac{a}{R}\right)=\lim_{\Lambda\to\infty}
\sum_{l}^{\Lambda R}\frac{2l+1}{2l(l+1)-x}-\log(\Lambda R)  
\end{equation}
If we identify the cutoff with some maximum angular momentum $\lambda$ via $\lambda(\lambda+1)\equiv(\Lambda R)^2$, then our expression above can be written as
\begin{equation}\label{eqn:blah blah}
\log\left(\frac{a}{R}\right)=\lim_{\lambda\to\infty}
\sum_{l}^{\lambda}\frac{2l+1}{2l(l+1)-x}-\frac{1}{2}\log(\lambda(\lambda+1))  
\end{equation}
The sum can be explicitly expressed in terms of the digamma function $\psi(x) = \frac{\textrm{d}}{\textrm{d}x} \log(\Gamma(x))$,
\begin{multline}
 \sum_{l}^{\lambda}\frac{2l+1}{2l(l+1)-x}=  \frac{1}{2} \left(\psi\left(\lambda -\frac{1}{2} \sqrt{2
   x+1}+\frac{3}{2}\right)+\psi\left(\lambda +\frac{1}{2} \sqrt{2
   x+1}+\frac{3}{2}\right)\right.\\
   -\left.\psi\left(\frac{1}{2}\left(1-\sqrt{2x+1} \right)\right) -\psi \left(\frac{1}{2} \left(1+\sqrt{2x+1}	\right) \right)\right)\ .
\end{multline}
In the limit $\lambda\to\infty$ the first two terms on the RHS above exactly cancel the logarithm term in Eq.~\eqref{eqn:blah blah}  What remains gives us our closed-form expression,
\begin{align}\label{eqn:luescher_0}
	\log\left( \frac{a}{R}\right) &= -\frac{1}{2} \left[\psi\left(\frac{1}{2}\left(1-\sqrt{2x+1} \right)\right) +\psi \left(\frac{1}{2} \left(1+\sqrt{2x+1}	\right) \right)	\right]\\
	&\equiv Z_0(x)\ .
\end{align}

\subsubsection{$L=1$}
The triangle inequality in this case requires that, given $l_1\equiv l$, the sum over $l_2$ is restricted to the values $|l-1|$, $l$, and $l+1$. However, the 3-$j$ symbol vanishes for $l_1=l_2=l$ (when $L=1$), and so Eq.~\eqref{eqn:tada} becomes the sum over two expressions only,
\begin{multline}
    \log\left(\frac{a}{R}\right)=%\lim_{\Lambda\to\infty}
    \frac{3}{2-x}\begin{pmatrix}
0 & 1 & 1\\ 0 & 0 & 0
\end{pmatrix}^2\\
+
\sum_{l\ge 1}^{\Lambda R}\left(\frac{4l^2-1}{2l^2-x}\begin{pmatrix}
l & l-1 &1\\ 0 & 0 & 0
\end{pmatrix}^2+\frac{(2l+1)(2l+3)}{2(l+1)^2-x}\begin{pmatrix}
l & l+1 &1\\ 0 & 0 & 0
\end{pmatrix}^2\right)\\
-\log(\Lambda R)\ .
\end{multline}
The first term on the RHS above comes from the $l=0$ contribution.  After simplifying the 3-$j$ symbols the sums can be performed and analytically expressed in terms of digamma functions.  The $\Lambda\to\infty$ limit can be subsequently taken, giving
\begin{align}\label{eqn:luescher_1}
	\log\left(\frac{a}{R}\right) &= -\frac{1}{2} \left[ \psi \left(1-\sqrt{\frac{x}{2}}\right)  + \psi \left(1+\sqrt{\frac{x}{2}}\right)\right]\\
	&\equiv Z_1(x)\ .
\end{align}
The fact that there exists a non-trivial quantization condition for $L=1$, despite the interaction being a pure contact interaction, comes from the fact that our general expression in Eq.~\eqref{eqn:tada} is derived using single-particle coordinates as opposed to relative coordinates.

\subsubsection{$L=2$}
The steps used for the $L=0,1$ cases can be analogously applied to $L=2$ (and higher).  Clearly the sum over $l_2$ for a given $l_1$ becomes more involved as $L$ becomes larger, and as such, the expressions become more complicated and cumbersome to express.  Therefore we do not show these steps here but instead provide the expression for $L=2$ without derivation:     
\begin{align}\label{eqn:luescher_2}
	\log\left(\frac{a}{R}\right) &= \frac{1}{12-8x} \left[2+3(x-2) \left\{\psi\left(\frac{1}{2} \left(3-\sqrt{-3+2x}\right)\right) + \psi\left(\frac{1}{2}\left(3+\sqrt{-3+2x}\right)\right) \right\}\right. \nonumber\\ 
	&\quad\quad+\, x \left. \left\{ \psi\left(\frac{1}{2} \left(1-\sqrt{1+2x}\right)\right) + \psi\left(\frac{1}{2} \left(1+\sqrt{1+2x}\right)\right) \right\}\right] \\
	&\equiv Z_2(x)\ .
\end{align}
\subsection{Limits and zeros of the quantization relations for $L=0,1$ cases}
\begin{figure}
    \centering
    \includegraphics[width=.5\textwidth]{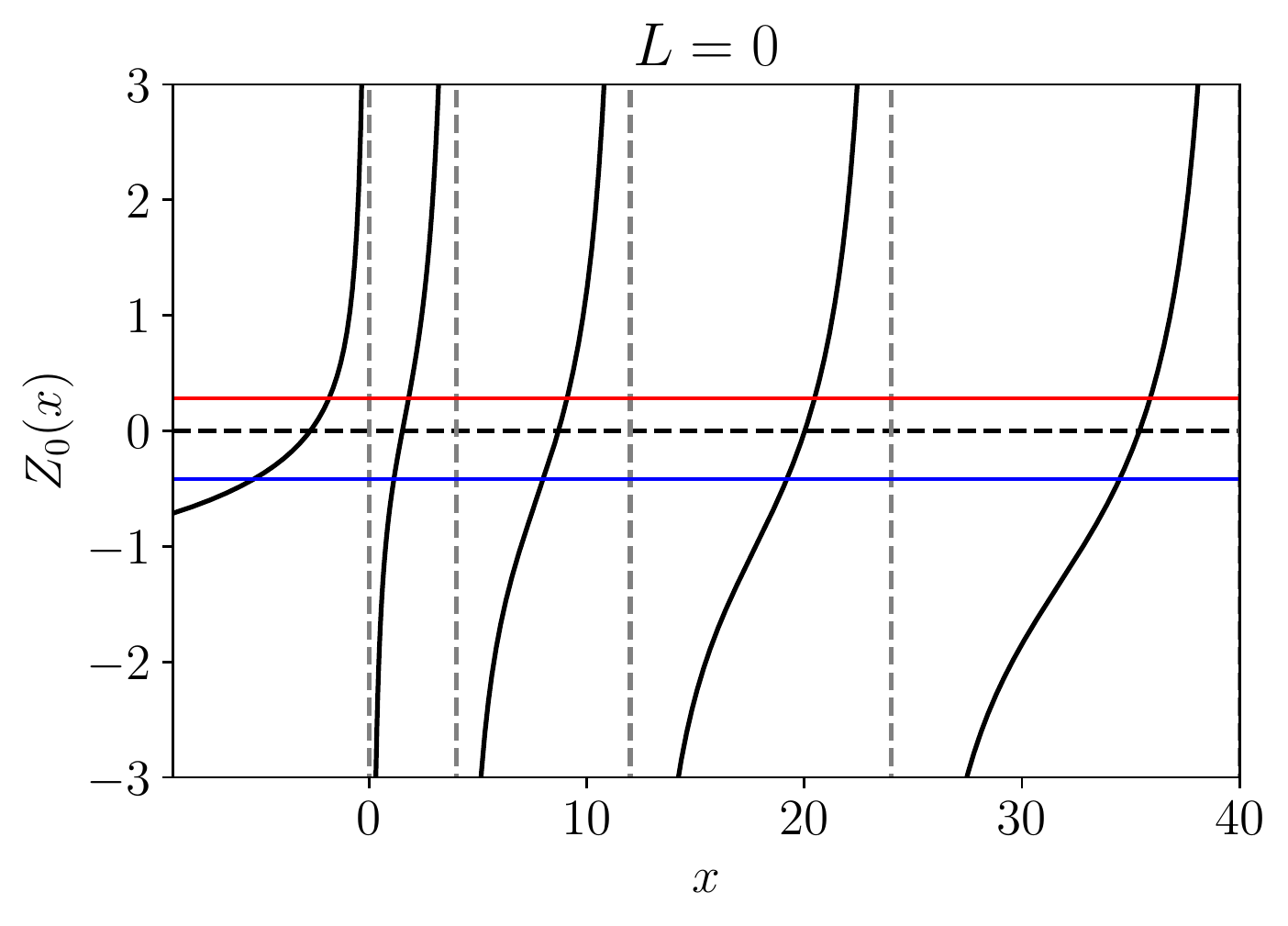}\includegraphics[width=.5\textwidth]{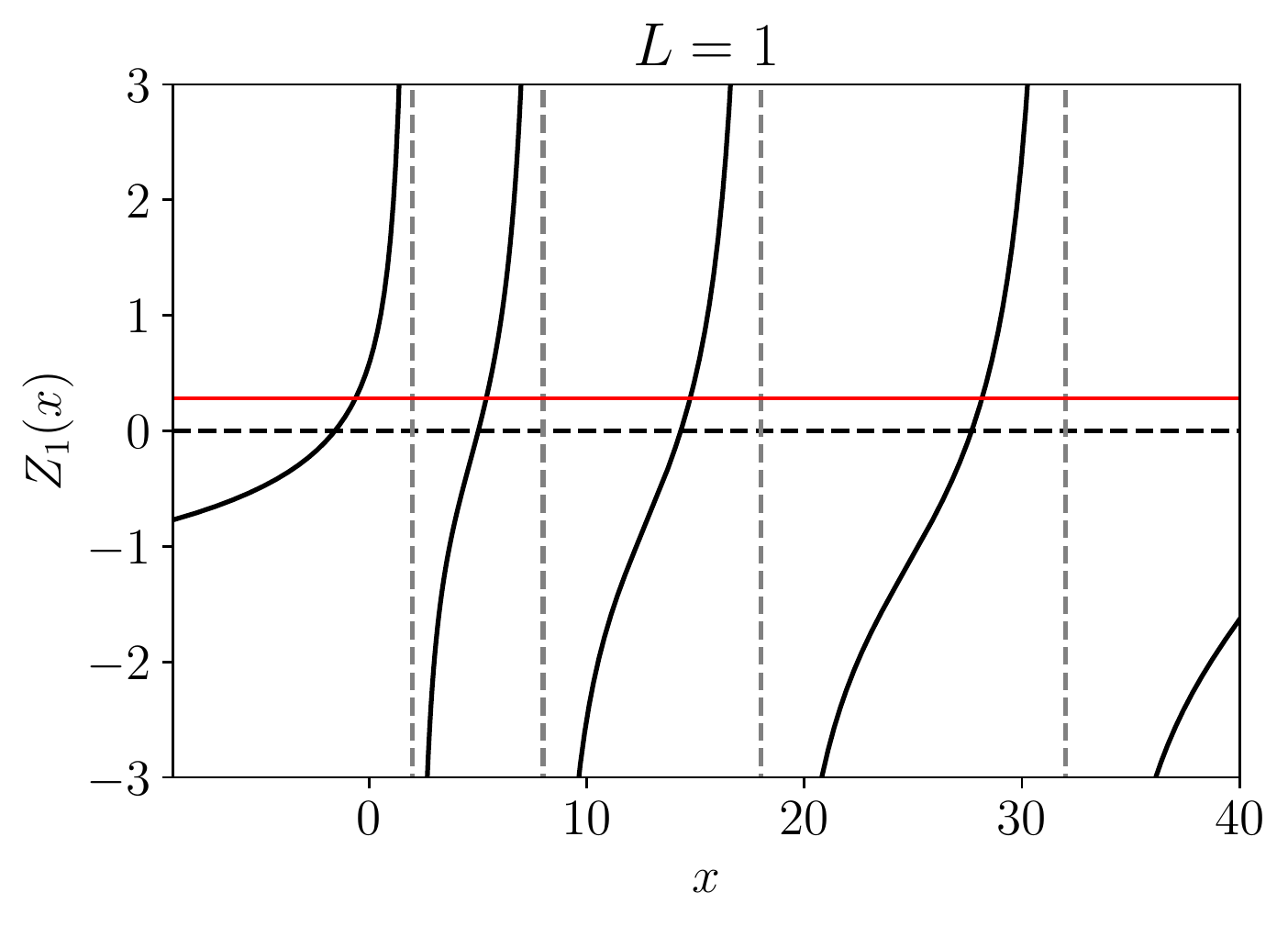}\\
    \includegraphics[width=.5\textwidth]{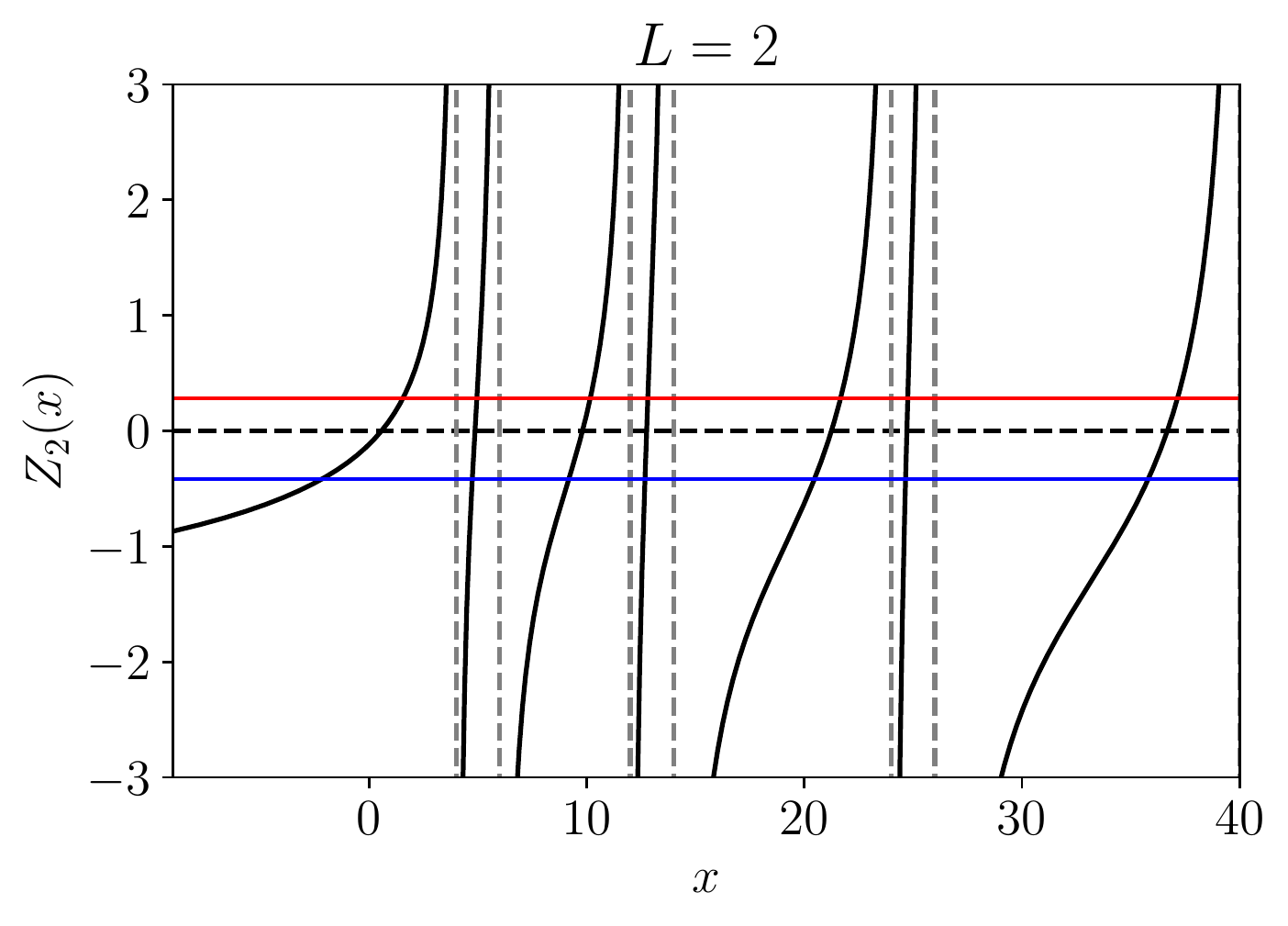}
    \caption{Quantization curves for the three lowest angular momenta $L$ given by eqs.~Eq.~\eqref{eqn:luescher_0},~Eq.~\eqref{eqn:luescher_1}, and~Eq.~\eqref{eqn:luescher_2}.  The vertical dashed lines correspond to the non-interacting energies.  The horizontal red line is the experimentally determined value of $-\frac{\pi}{2\tilde a}$ in the spin-singlet ($S=0$) case, while the blue line corresponds to the spin-triplet ($S=1$) case, both of which are described and used in Sect.~\ref{sec:halo nuclei}}
    \label{fig:curves}
\end{figure}

The structure of these quantization equations for energies $x \in \{-9, 40\}$ is displayed in fig.~\ref{fig:curves}. These will be utilized to find solutions to the Schr\"odinger equation for two particles on a sphere in the following chapter. One may notice the divergent parts of each graph, which corresponds to the case of no interaction.
%\begin{figure}
%    \centering
%    \includegraphics[width=0.6\textwidth]{curves.pdf}
%    \caption{L\"uscher curves of the three lowest rotational bands.}
%    \label{fig:curves}
%\end{figure}

As already mentioned earlier, in two dimensions the scattering length $a\ge0$ \cite{Pupyshev:2014xua}.  In the limit $a\gg R$, the solutions to Eq.~\eqref{eqn:luescher_0} approach the non-interacting energies from below.  In the limit $a\ll R$ we have, in addition to the deeply bound solution $x\to-\infty$ (i.e. the so-called ``dimer solution"), solutions that also approach the non-interacting energies, but now from above. We can expand the solutions $x$ about the non-interacting energies by considering the limit $|\log(a/R)|\gg1$. For the $n^{\rm{th}}$ solution, where $n\in \mathbb{Z}_{\ge 0}$, we find for the $L=0$ case
\begin{equation}\label{eqn:L0 limit}
    x=2n(n+1)-\frac{2n+1}{\log(a/R)}%+\frac{\frac{1}{2}-\gamma_E}{\log(a/R)^2}
    +\mathcal{O}\left(\log(a/R)^{-2}\right)\ .
\end{equation}
This expression is valid for both limits $a\to\infty$ and $a\to 0$ (keeping $R$ fixed).  The bound dimer solution valid as $a\to 0$ scales as
\begin{equation}\label{eqn:dimer limit}
    x=-\frac{2R^2}{a^2}+\mathcal{O}(a^2)\ .
\end{equation}
Given that $x=2mER^2$, this corresponds to the standard dimer binding energy $E=-\frac{1}{m a^2}$.  
Similarly, for $L=1$ we have
\begin{equation}
    x=2(n+1)^2-\frac{2(n+1)}{\log(a/R)}%+\frac{\frac{1}{2}-\gamma_E}{\log(a/R)^2}
    +\mathcal{O}\left(\log(a/R)^{-2}\right)\ ,
\end{equation}
which is again valid for both limits $a\to\infty$ and $a\to 0$ while keeping $R$ fixed.  The dimer solution  scales identically the same as the $L=0$ case, Eq.~\eqref{eqn:dimer limit}.
\begin{table}[t]
    \centering
    \begin{tabular}{|c|c|c|}
        \hline
          L=0 & L=1 & L=2 \\
          \hline
         \SI{-2.69519416311127 \pm 0.00000000000001} & \SI{-1.56227783993538 \pm 0.00000000000001} & \SI{0.57785048142503 \pm 0.00000000000001}{} \\
         \SI{1.53660948605491 \pm 0.00000000000001} & \SI{5.02284537252901 \pm 0.00000000000001} & \SI{4.86959138770876 \pm 0.00000000000001}{}\\
         \SI{8.70562260382481 \pm 0.00000000000001} & \SI{14.32769206430169 \pm 0.00000000000001} & \SI{12.76394742135846 \pm 0.00000000000001}{}\\
         \SI{20.02549569293160 \pm 0.00000000000001} & \SI{27.69206196072471 \pm 0.00000000000001} & \SI{21.24994781781113 \pm 0.00000000000001}{}\\
         \vdots & \vdots & \vdots\\
         \hline
    \end{tabular}
    \caption{Zeros $x_0$ for the three lowest rotational bands.}
    \label{tab:zeros}
\end{table}

Another interesting limit is to consider the case when $a/R=1$, corresponding to the $|\tilde a|\to \infty$ limit.\footnote{If both $a\to\infty$ and $R\to\infty$, while $a/R=1$, then this corresponds to the 2-d unitary limit where all length scales have been ``integrated out".}  Solutions to Eq.~\eqref{eqn:tada} in this case occur when the curves in Fig.~\ref{fig:curves} intersect the x-axis, corresponding to zeros of the quantization equations.  We provide these zeros to machine precision for $L=0$, $1$, and $2$ in \autoref{tab:zeros}.  For both $L=0$ and $L=1$ there exists an $x_0<0$ solution corresponding to a bound state in this limit. 

\begin{figure}
    \centering
    \includegraphics[width=.8\textwidth]{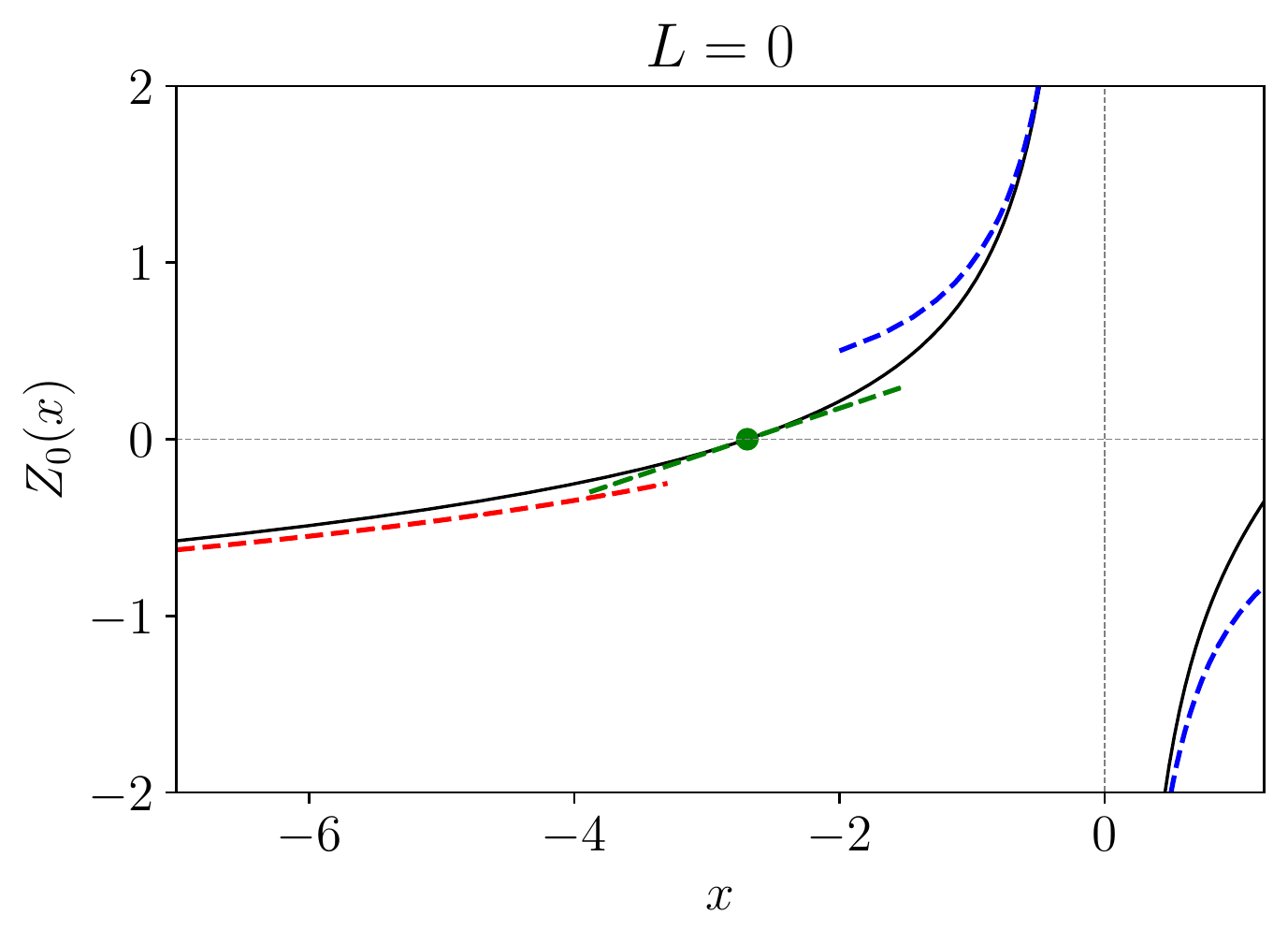}
    \caption{The limits of the quantization equation for the $L=0$ case.  The red dashed curve corresponds to Eq.~\eqref{eqn:dimer limit}, the blue dashed line Eq.~\eqref{eqn:L0 limit}, and the green dashed curve Eq.~\eqref{eqn:zero limit} using $x_0=-2.69519$.  The solid black line is given by Eq.~\eqref{eqn:luescher_0}.} 
    \label{fig:L0 limits}
\end{figure}

The behavior of the eigenvalue solutions near a general zero $x_0$ when $a\approx R$ for the $L=0$ band is
\begin{equation}\label{eqn:zero limit}
    x=x_0+
      \frac{4 \sqrt{2 x_0+1}\log\left(a/R\right)}{\psi ^{(1)}\left(\frac{1}{2}-\frac{1}{2} \sqrt{2 x_0+1}\right)-\psi
   ^{(1)}\left(\frac{1}{2}+\frac{1}{2} \sqrt{2 x_0+1}\right)}
    +\mathcal{O}\left(\log(a/R)^2\right)\ ,
\end{equation}
while for $L=1$ it is
\begin{equation}
    x=x_0+
   \frac{4 \sqrt{2x_0} \log\left(a/R\right)}{\psi ^{(1)}\left(1-\sqrt{\frac{x_0}{2}}\right)-\psi
   ^{(1)}\left(1+\sqrt{\frac{x_0}{2}}\right)}
    +\mathcal{O}\left(\log(a/R)^2\right)\ ,
\end{equation}
where $\psi ^{(1)}(z)\equiv \frac{d}{dz}\psi(z)$. In Fig.~\ref{fig:L0 limits} we plot these limiting expressions and compare them to the exact solution for the $L=0$ case.

\subsection{Comparison with $S^1\times S^1$ topology and 2-D harmonic oscillator}
As mentioned earlier, the quantization condition for two particles interacting in a confined space has been determined in other 2-D systems.  Here we take the opportunity to compare our $L=0$ result Eq.~\eqref{eqn:luescher_0} with its analog in the $S^1\times S^1$ geometry and the harmonic oscillator.  

Busch \emph{et al.}\cite{Busch:1998} have derived the case for the 2-D harmonic oscillator with frequency $\omega$,
\begin{equation}\label{eqn:HO}
    \log\left(\frac{a}{b}\right)=-\frac{1}{2}\psi\left(\frac{1}{2}-\frac{x}{2}\right)\ ,
\end{equation}
where $x=E/\omega$ with $E$ the eigenenergy and $b=1/\sqrt{2m\omega}$ is the oscillator parameter.  

For a 2-D square lattice of side $L$ with periodic boundary conditions (i.e. the torus or $S^1\times S^1$ topology), a thorough derivation is provided in \cite{Beane:2010ny}, giving
\begin{equation}\label{eqn:S1S1}
   \frac{2}{\pi} \log\left(2\pi\frac{a}{L}\right)=\lim_{\Lambda\to\infty}\frac{1}{\pi^2}\sum_{\bm{n}}^{|\bm{n}|\le\Lambda}
    \frac{1}{\bm n^2-x}-\frac{2}{\pi}\log\left(\Lambda\right)\equiv \frac{1}{\pi^2}S_2(x)\ .
\end{equation}
Here $S_2(x)$ is the two dimensional zeta function, $\bm n = (n_i,n_j)\ \in\mathbb{Z}^2$, and $x=m E L^2/(4\pi^2)$.  
\begin{figure}
    \centering
    \includegraphics[width=.5\textwidth]{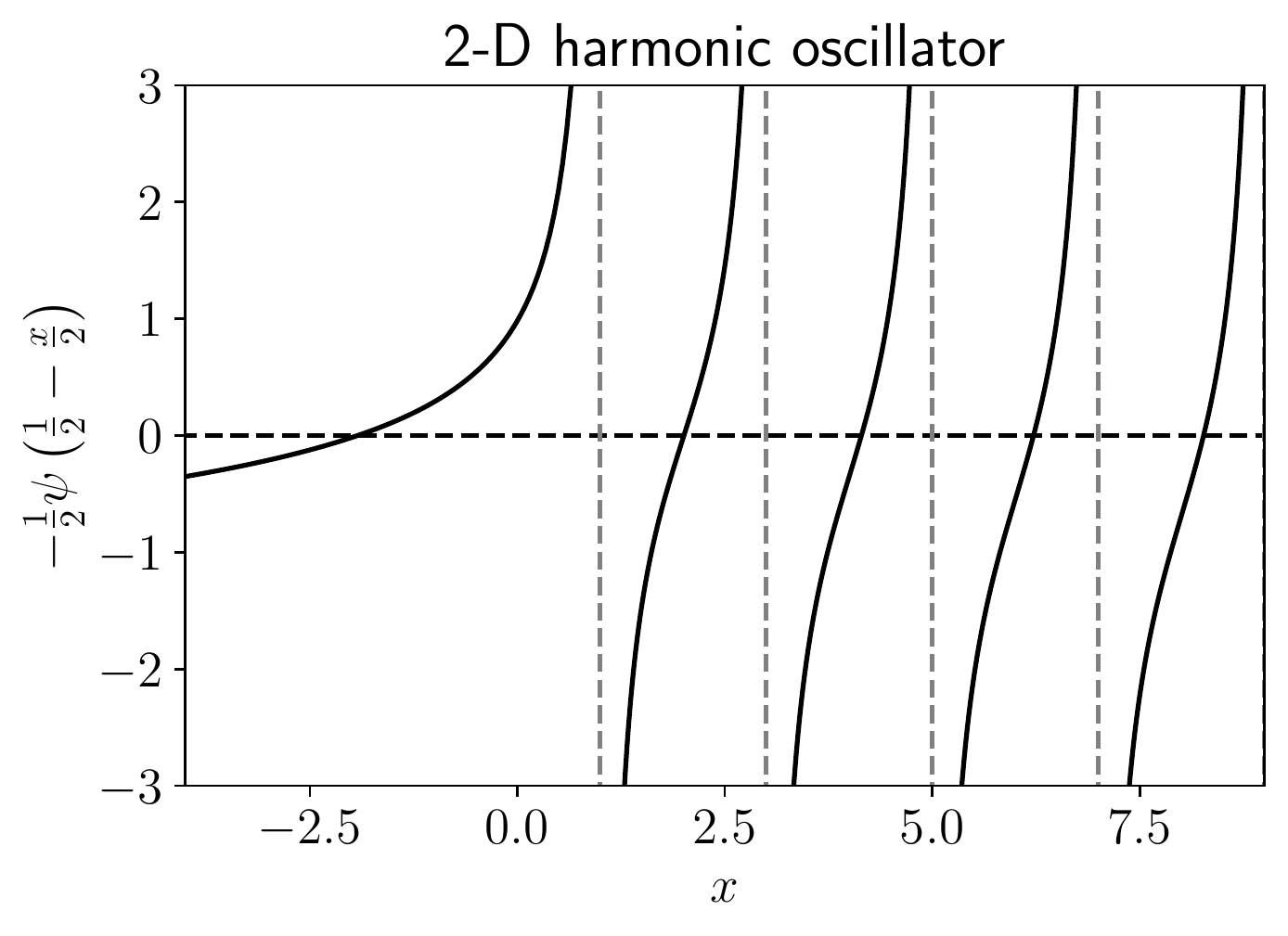}\includegraphics[width=.5\textwidth]{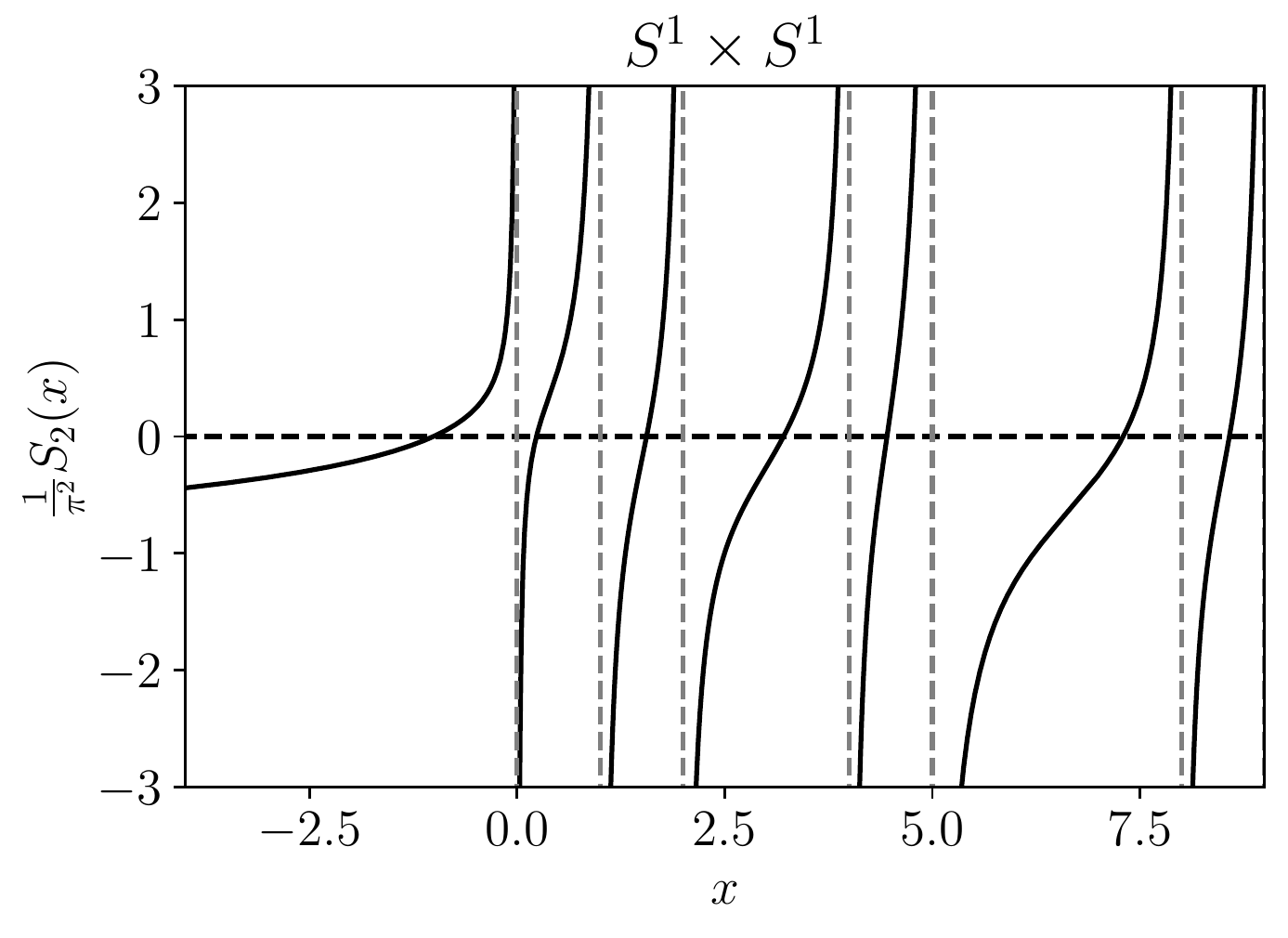}
    \caption{S-wave quantization conditions for the 2-D harmonic oscillator (left, Eq.~\eqref{eqn:HO}) and the $S^1\times S^1$ topology (right, Eq.~\eqref{eqn:S1S1}).  The vertical dashed gray lines correspond to the non-interacting energies of each system.  Compare with the $L=0$ quantization relation on $S^2$ (top panel of Fig.~\ref{fig:curves} and Eq.~\eqref{eqn:luescher_0}) }
    \label{fig:ho and S1S1} 
\end{figure}
The dependence of these functions on $x$ is shown Fig.~\ref{fig:ho and S1S1}, and should be compared with the top panel of Fig.~\ref{fig:curves}.  In all cases, the curves approach their respective non-interacting solutions in both limits $a\to\infty$ and $a\to 0$, all other parameters held fixed.  Furthermore, all cases have $x_0<0$ as the lowest x-axis intercept, corresponding to a bound state in the limit $a/R=a/b=2\pi a/L=1$.

Because the harmonic oscillator and torus results are derived in geometries in which the CM and relative coordinates are exactly separable, it is trivial to determine their quantization conditions for \emph{non-zero CM motion}.  Such motion corresponds to a quantized CM energy $E_{CM}$, which in dimensionless units is $x_{CM}=E_{CM}/\omega$ for the harmonic oscillator and $x_{CM}=mE_{CM}L^2/(4\pi^2)$ for the torus.  To obtain the quantization curves for these cases, one shifts the curves in Fig.~\ref{fig:ho and S1S1} to the right by exactly $x_{CM}$.  In this manner any non-zero CM quantization curve in the harmonic oscillator and torus cases can be obtained directly from the curves shown in Fig.~\ref{fig:ho and S1S1}.

For $S^2$ there is no trivial connection to the lowest energy quantization curve given by $L=0$, since the geometry is not amenable to CM and relative motion separability.  The analog of different CM motion manifests itself as different rotational bands $L$, and as can be seen from Fig.~\ref{fig:curves} the $L=1$ and $L=2$ curves (and in general $L>2$) are not connected to $L=0$ by any constant shift in the $x$-axis. 

\section{\label{sec:halo nuclei}Application: predicting energy levels of two-nucleon halo nuclei}
Halo nuclei consist of a tightly bound core of nucleons surrounded by small group of loosely bound, or halo, nucleons.  The resulting nuclei appear much larger than the radius of the original tightly bound core.
A classic example is the $^{11}$Li halo nucleus originally found by I. Tanihata \emph{et al.} \cite{Tanihata:1985psr}.  This nucleus can be decomposed into a three-body system, $^9$Li + 2n, where the $^9$Li constitutes the tightly bound core and the two neutrons the halo nucleons that are considered to be loosely bound and interacting.  Another example is $^6$He, which can also be decomposed into a tightly bound core, $^4$He, plus two halo neutrons, again loosely bound and interacting.  Both of these systems are only stable\footnote{Stable in this context applies only to the strong interaction.} as a three-body constellation, and therefore are considered \emph{borromean} \cite{Zhukov:1993aw}.

A simple, albeit crude, approximation to these systems is to assume that the two nucleons are constrained to interact on a sphere with halo radius $R$ and that the core is located at the center of this sphere. The confinement of the halo nucleons is assumed to be due to some non-trivial interaction with the core, which we approximate as infinitely massive and therefore non-dynamical.\footnote{Such an approximation has been used to describe doubly-excited atomic electrons interacting via a contact interaction\cite{PhysRevA.25.1513,PhysRevA.28.1974} and via a modified coulomb interaction\cite{PhysRevA.75.062506}, for example.}  When the core has its own angular quantum numbers, we may couple the angular momentum of the halo nucleons with that of its core, but aside from that, the core has no other influence on the halo nucleons.  If we further assume that the interaction between the nucleons is contact in nature, then our formalism of the previous section directly describes this situation.\footnote{Naturally there exist more sophisticated models and calculations of these systems, see e.g. Refs.\cite{Caprio:2014iha,Hammer:2017tjm,Hongo:2022sdr} and references within.}  Under this approximation radial excitations are not possible and therefore there are only vibrational excitations for each rotational band.

Nucleons are of course fermions with spin and isospin equal to $1/2$.  To incorporate our results from the previous section, we must take the nucleons' spins, isospins, and their Pauli-exclusion into account.  For the two nucleons to `feel' the s-wave interaction, we must couple their spins and isospins to total spin and isospin $S=0$, $T=1$ (e.g. `spin-singlet' two-neutron system) or $S=1$, $T=0$ (i.e. `spin-triplet' deuteron system), respectively.  We then couple their total spin $S$ and angular momentum $L$ to make total angular momentum $J_{\rm NN}$.  An anti-symmetric two-nucleon wavefunction requires 
\begin{displaymath}
L+S+T-l_1-l_2={\rm odd}\ ,
\end{displaymath}
and this in turn restricts the allowed angular momentum $L$ of the two nucleons. The total angular momentum $J_{\rm NN}$ of the halo nucleons is then coupled with the angular momentum of the core to obtain the total angular momentum of the halo system $J$. Finally, the parity of the two-nucleon system is
\begin{displaymath}
(-1)^L\ ,
\end{displaymath} 
and is multiplied with the parity of the core to obtain the overall parity $\pi$ of the halo system.

Before we can use our formalism to predict energy levels, however, we have to tune the parameters (i.e. $\tilde a$ or equivalently $a/R$) of our theory.  We now describe in detail how we use the low-energy spectrum of the $^6$He and $^{11}$Li halo nuclei to determine these parameters.  In particular, these systems will allow us to determine the spin-singlet scattering length $\tilde a_0$.  We also consider the $^6$Li system which will allow us to determine the spin-triplet scattering length $\tilde a_1$.

\subsection{Helium-6}
\begin{figure}
    \centering
    \includegraphics[width=.5\textwidth]{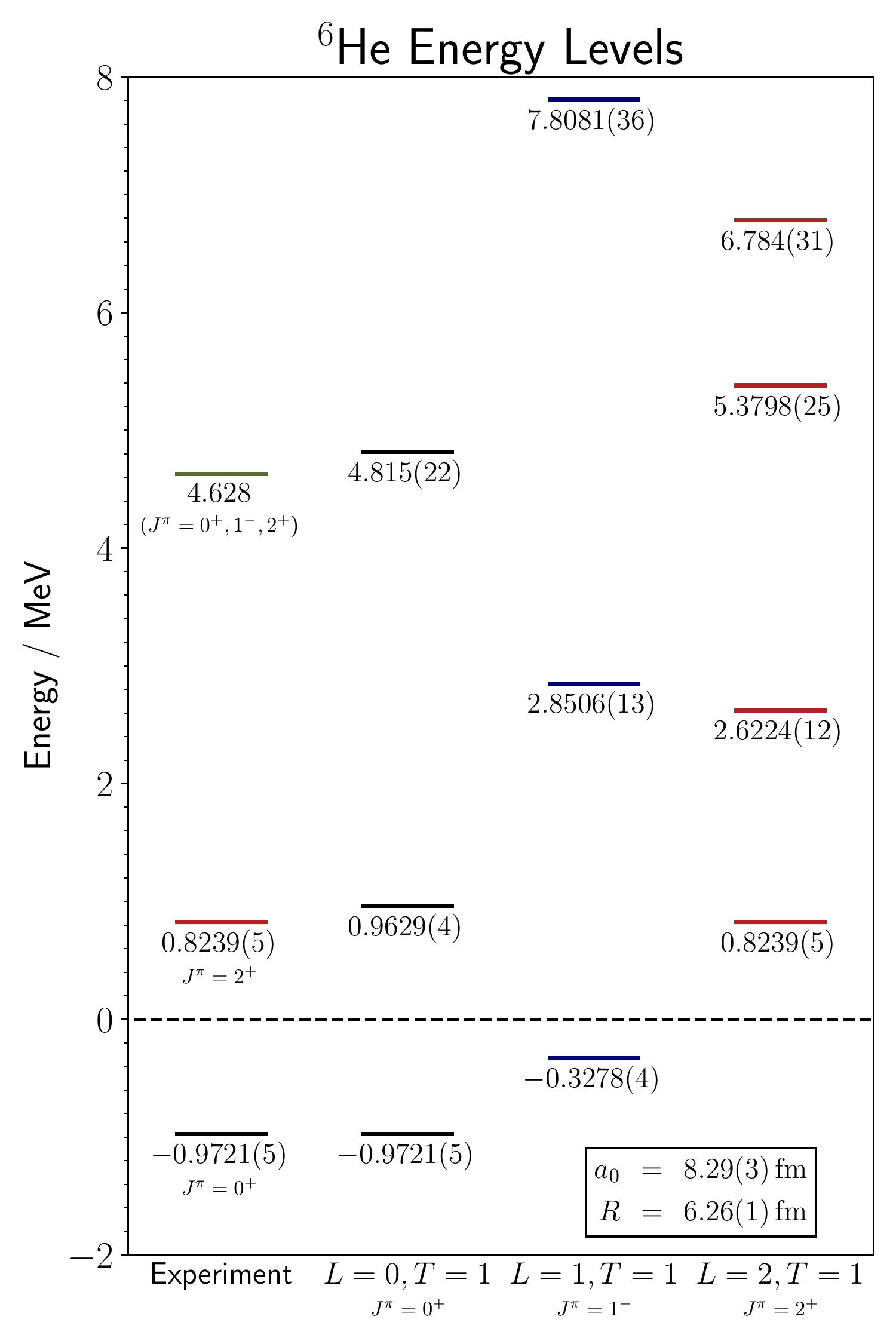}\includegraphics[width=.5\textwidth]{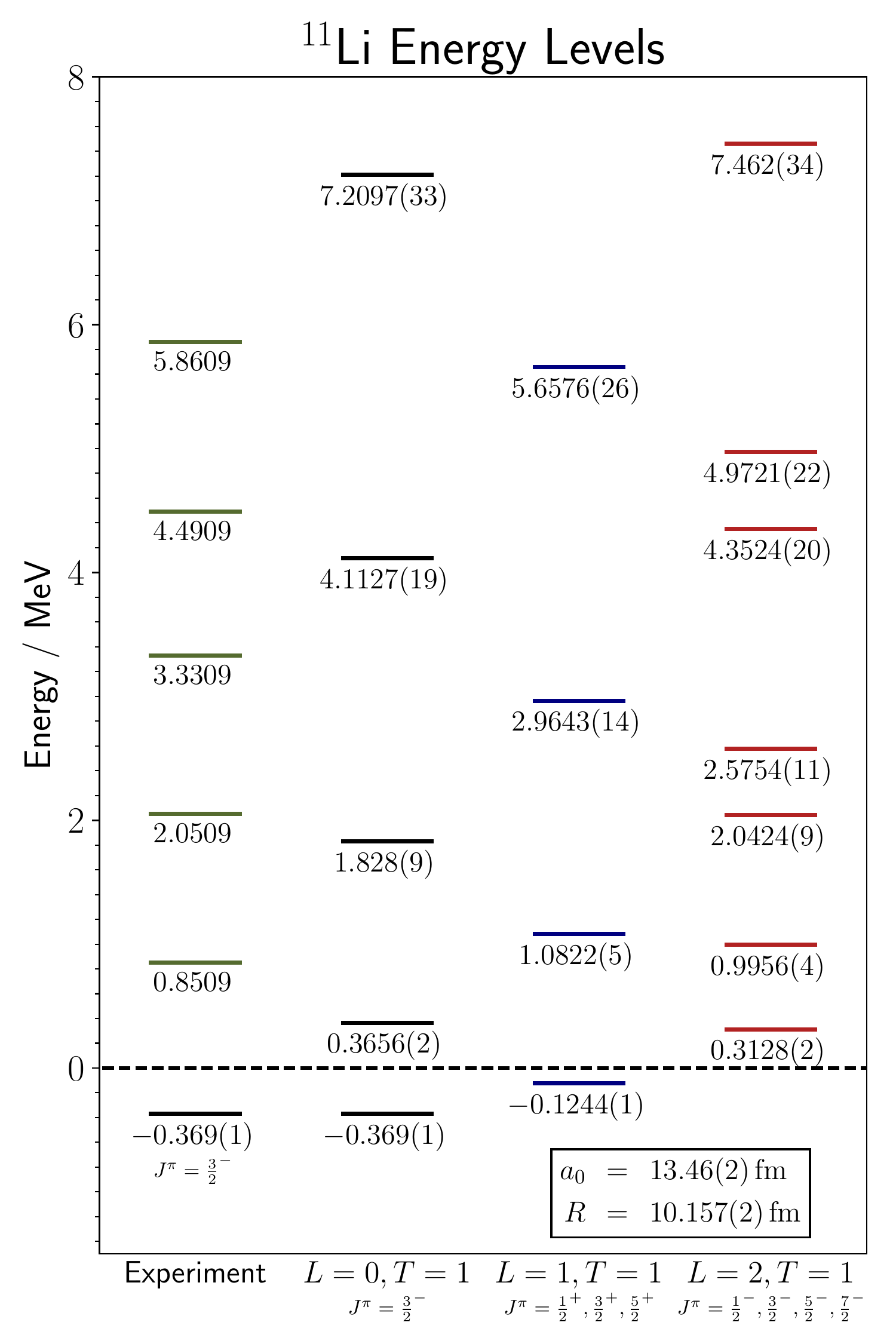}
    \caption{Two-neutron halo energy levels for select rotational bands of $^6$He (left) and $^{11}$Li (right), compared to experiment. The lowest two experimental energies of $^6$He and the lowest experimental energy of $^{11}$Li were used to determine $\tilde a_0$ and our model parameters shown in the boxed insets.  These were then used to make the predictions of the energy levels.  Where possible we have provided the $J^\pi$ quantum numbers of the levels, and color-coded the levels to match the quantum numbers.  The uncertainties do not represent widths of the levels, but rather are the uncertainties of our model predictions, given the level of accuracy of the experiments.}
    \label{fig:he6 li11 energy levels}
\end{figure}

Here we have two neutrons surrounding a $^4$He core.  The two neutrons are thus in the $S=0$, $T=1$ channel. The three lowest allowed angular momentum bands are $L=0$, $L=1$, and $L=2$, with $L=1$ being odd in parity and the others even.  As the $^4$He core has $J_{\rm C}^{\pi_{\rm C}}=0^+$ angular momentum, the total angular momenta of the halo nucleus for these bands are simply $J^\pi=J_{\rm NN}^\pi=0^+,\ 1^-$, and $2^+$. Within our approximation the interaction of two neutrons on a 2d surface is described solely by the parameter $\tilde a_{0}$, where we add the subscript $0$ to denote that this parameter is for the spin-singlet $S=0$ system.  This parameter is \emph{independent} of the halo nucleus.   As it is also dimensionless, a single empirical (dimensionful) energy is not sufficient to constrain this parameter and therefore a second energy is required.  We use the experimental $J^\pi=0^+$ and $2^+$ energies of the $^6$He halo nucleus~\cite{He6_data,Tilley:2002vg}, measured relative to the $^4{\rm He} +n+n$ threshold,  to constrain the dimensionful parameters $a_0$ and $R$ of our theory, which we stress are halo nucleus \emph{dependent}. We then obtain $\tilde a_0$ by the relation~Eq.~\eqref{eqn:reduced scattering length}.  We find 
\begin{equation}\label{eqn:a0}
    \tilde a_0 = \SI{-5.58\pm 0.06}\ .
\end{equation}
The experimental energies used to obtain this value, as well as the resulting $a_0$, $R$, and predicted energy levels of our model for the $J^\pi=0^+,\ 1^-$ and $2^+$ rotational bands, are given in Fig.~\ref{fig:he6 li11 energy levels}.  We take the mass of the neutron as $m=939.565$ MeV.

To obtain the errors of the fit parameters quoted in Fig.~\ref{fig:he6 li11 energy levels}, we first assume that the experimental errors for the $J=0$ and $J=2$ energies are uncorrelated and follow a normal distribution with width dictated by their respective errors.  We then sample these energies from their distributions, each time performing our fit to obtain $a_0$, $R$, and $\tilde a_0$, and we tally these results.  The mean of these tallies is our quoted values of these terms in Fig.~\ref{fig:he6 li11 energy levels}, and the standard deviation their errors.
%\footnote{Note that the errors for $a$ and $R$ obtained in this manner are \emph{correlated}, and therefore the quantities $a/R$ and $\log\left(a/R\right)$ tend to have a smaller error then $a$ or $R$ alone.}.  
Our sample size is 10,000.

The determined value of $\tilde a_0$ then fixes $\log\left(a_0/R\right)$ through the relation~Eq.~\eqref{eqn:reduced scattering length}, which we show as the red line in the $L=0,\ 1$ and $2$ plots in Fig.~\ref{fig:curves}.  The intercept of this red line with the solid black curves in these plots gives us our energy solutions.  Our fitting procedure is guaranteed to reproduce the lowest $0^+$ and $2^+$ experimental energies and their errors, as these were used to obtain our fit parameters.  The higher intercepts then provide our predicted energy levels shown in Fig.~\ref{fig:he6 li11 energy levels}.

As already mentioned above, the applicability of our model is quite limited due to its extreme simplicity, and this is quite obvious when looking at its predicted $J^\pi=1^-$ energies.  Our model predicts as its lowest state a \emph{negative} energy solution, although experimentally no such state exists.  Furthermore, there exist positive energy solutions that are predicted in other rotational bands that have no obvious experimental counterparts.  It is also interesting to compare our estimate of the halo radius $R=6.258(15)$ fm which is nearly a factor of two larger than the experimental result of $R_{\rm exp}=3.08(10)$ fm \cite{halo_radii_1}.   Again, this disagreement is not surprising given the level of crudeness of our model.  %Furthermore, the difference between our calculated result and experiment is to a certain extent arbitrary, as we are free to choose the length scale in Eq.~\eqref{eqn:reduced scattering length}.  If we instead had chosen the \emph{diameter} of the sphere as our length scale, then our calculated radius $R=3.13(1)$ fm would be much closer to experiment.  However, in this case our bound energy would scale as $E\sim \frac{-8}{ma^2}$ which disagrees with the scaling of the standard dimer energy.  We stress, however, that regardless of the chosen length scale, our predicted energy levels would be the same as those shown in \autoref{tab:energy_levels}. {\bf This is hard to understand, right?}

\subsection{Lithium-11}
The $^9$Li core has angular quantum numbers $3/2^-$, and for the $^{11}$Li halo system there is only the measured $J=3/2^-$ ground state energy $E_0 = \SI{-0.369}{MeV}$ \cite{li11,Kelley:2012qua} that has definitive quantum numbers assigned. However, given that we determined $\tilde a_0$ in the previous section (which in our approximation is independent of halo nucleus), we have sufficient information to determine $a_0$ and $R$ for this system.  In this case we tally fit results obtained from uncorrelated samplings of $E_0$ and $\tilde a_0$ to arrive at $a_0$ and $R$, and then subsequently predict the higher energy levels.  Our results are given in the right panel of Fig.~\ref{fig:he6 li11 energy levels}.  When coupling the angular momentum $J_{NN}$ of the halo nucleons with that of the $3/2^-$ core, our model predicts multiplets of energies in the $L=1$ and $L=2$ cases.  We label these multiplets in our figure. 

As in the $^6$He case, our model predicts another negative energy solution near threshold coming from the $L=1$ case, which is not seen experimentally.  Our model again predicts many positive energy solutions that have no obvious experimental counterparts.  We compare our estimate of the halo radius $R=10.154(24)$ with its experimental value, $R_{\rm exp}=\SI{6.5 \pm 0.3}{fm}$ \cite{halo_radii_1}.  Again, this level of disagreement is not surprising given the simplicity of our model.

\subsection{Lithium-6}
Though not technically considered a halo nucleus, the small separation energy for $^4{\rm He}+n+p$  breakup (small compared to the binding energy of its $^4$He core) suggests that the nucleus is extended in size.  We therefore assume that the $^4$He acts as the core and the `halo' nucleons for this system consist of a neutron and proton.  This system supports both $S=0,\ T=1$ and $S=1,\ T=0$ channels, therefore we expect the spectrum to be much richer than in the previous two examples.  Both of these channels can couple to $L=0$ and $2$ angular momenta for positive parity.  The $S=0,\ T=1$ channel can also couple with the negative parity $L=1$ angular momentum. When coupled with $J_C^{\pi_C}=0^+$ of the core, we have $J^\pi(T)=1^+(0),\ 2^+(0),\ 3^+(0),\ 0^+(1),\  2^+(1)$, and $1^-(1)$ as possible quantum numbers.  We use $m=2\mu=938.918$ MeV, where $\mu$ is the reduced mass of the proton and neutron.

\begin{figure}[h!]
    \centering
    \includegraphics[width=\textwidth]{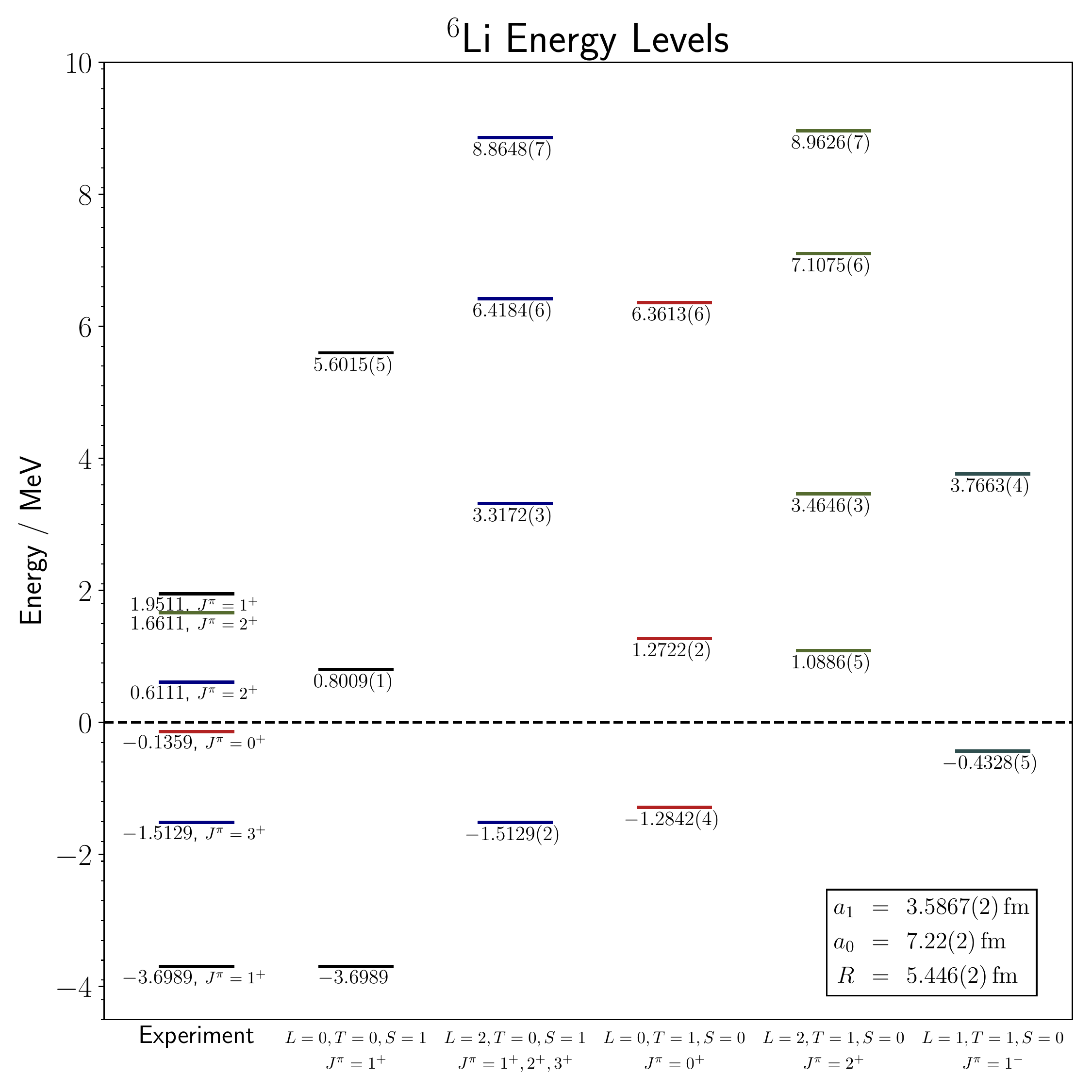}
    \caption{Two-nucleon halo energy levels for select rotational bands of $^6$Li and comparison with experimental values, where available~\cite{Tilley:2002vg}. The lowest two experimental energies are used to determine the parameters of our model, which are shown in the boxed inset.  Where possible we have provided the $J^\pi$ quantum numbers of the levels, and color-coded the levels to match the quantum numbers.  The uncertainties do not represent widths of the levels, but rather are the uncertainties of our model predictions.}
    \label{fig:Li6}
\end{figure}

We assume isospin charge symmetry, meaning that the (dimensionless) scattering length in the spin-singlet $S=0$ channel is the same as that determined in the $^6$He case given in Eq.~\eqref{eqn:a0}.  To determine the spin-triplet $S=1$ scattering length, we again use the two lowest experimental energies~\cite{Tilley:2002vg} of this system, measured relative to the $^4{\rm He}+n+p$ breakup threshold.  Here the two lowest energies have the quantum numbers $1^+$ and $3^+$ states. We find
\begin{equation}\label{eqn:a1}
    \tilde a_1 = \SI{3.760\pm0.007}\ ,
\end{equation}
where the subscript $1$ denotes the $S=1$ spin-triplet system.  Note the sign change compared to the spin-singlet case in~Eq.~\eqref{eqn:a0}.\footnote{A similar sign change occurs for the two-nucleon scattering lengths in \emph{three} dimensions.}  As before, the determined value of $\tilde a_1$, along with the experimental energies, fixes $\log\left(a_1/R\right)$ through the relation~Eq.~\eqref{eqn:reduced scattering length}. We show this result as the blue line in the $L=0$ and $2$ plots in Fig.~\ref{fig:curves}.  The intersection of this blue line with the black curves gives us our predicted energy levels.  The experimental energies, our determined $a_0$, $a_1$ and $R$ parameters for this system, as well as our predicted energy levels are shown in Fig.~\ref{fig:Li6}.

It is interesting to note that relative to the $^4{\rm He}+n+p$ breakup threshold, the $^6$Li system has three positive parity negative energy states.  The two lowest energies, both in the $T=0$ channel, are exactly captured in our model, since we use these energies to fit our parameters $\tilde a_1$ and the combination $a_1$ and $R$. But our model also predicts a third positive parity negative energy corresponding to the $T=1$ channel.  This is due to the spin-singlet scattering length $\tilde a_0$ and the energy of this state coincides with the intersection of the red line with the lowest black curve of $L=0$ in Fig.~\ref{fig:curves}. The correct ordering of these levels is captured in our model, however the experimental value of this state is much closer to the $^4{\rm He}+n+p$ threshold, while our prediction is significantly lower in energy.  Lastly, our model predicts a near threshold negative energy in the negative parity $J^\pi(T)=1^-(1)$ band due to the coupling with $L=1$, which is not observed experimentally.

\section{\label{sec:conclusion}Conclusion}
In this paper we derived the quantization condition for two-particles constrained to a sphere, or $S^2$, and under the assumption that they interact via a contact interaction.  We show how the energy levels of the system are related to the reduced scattering length $a$ and radius of the sphere $R$.  As the constraint on $S^2$ represents a non-inertial frame, the system is not amenable to a separation of CM and relative coordinates.  As such, there is an infinite tower of solutions for each total angular momentum $L$, each of which is different and \emph{not} related by any constant offset from each other.  We provide a solution for any $L$ in terms of a general sum, but for the $L=0$, $1$, and $2$ cases we obtain closed form expressions.  We also derive expressions in the limit $a\gg R$, $a\ll R$, and $a\to R$.  

We then applied our formalism to select two-nucleon halo nuclei under the assumption that the halo nucleons are confined to a sphere of halo radius $R$ and the core is infinitely massive and therefore non-dynamical.  We tuned our system-dependent parameters to the low-lying spectrum of these halo nuclei and determined the spin-singlet $S=0$ and spin-triplet $S=1$ scattering lengths.  We then used these results to predict the higher-lying spectrum.  Our results for the halo radius disagreed by up to a factor of two from experiment, but given the level of crudeness of our approximation this was not a surprising result.    

Finally, our formalism, and its application to two-nucleon halo nuclei, provides another excellent pedagogical example of two quantum mechanical interacting particles, but this time within a non-inertial frame.

\begin{acknowledgments}
We thank Christoph Hanhart for getting us started on this project, and Evan Berkowitz, Johann Ostmeyer, and Ulf Mei{\ss}ner for their critical reading of the manuscript.  We are also immensely indebted to Andreas Nogga for his insightful comments. This work was supported in part by the NSFC and the Deutsche Forschungsgemeinschaft (DFG, German Research
Foundation) through the funds provided to the Sino-German Collaborative
Research Center TRR110 “Symmetries and the Emergence of Structure in QCD”
(NSFC Grant No. 12070131001, DFG Project-ID 196253076 - TRR 110). 
\end{acknowledgments}

\appendix

\section{\label{sec:matrix element}Derivation of $\Braket{(l_1l_2)LM|\hat{V}_{12}|(l_1'l_2')LM}$}

In Eq.~\eqref{eq:matrix_element} we stated the result of the matrix element $\Braket{(l_1l_2)LM|\hat{V}_{12}|(l_1'l_2')LM}$ which we will now go over in more detail. We start by inserting two complete sets of position eigenstates in order to evaluate the potential $\hat{V_{12}}$ in coordinate space and furthermore make use spherical harmonics.
\begin{multline}\label{eqn:ME}
\langle l_1m_1;l_2m_2|\hat V_{12}
|\lambda_1\mu_1;\lambda_2\mu_2\rangle
=\\
\frac{C(\Lambda)}{R^2}\int d\hat{\bm r}_1d\hat{\bm r}_2\langle l_1m_1;l_2m_2|\hat{\bm r}_1;\hat{\bm r}_2\rangle
\delta\left(\hat {\bm r}_1-\hat{\bm r}_2\right)
\langle \hat{\bm r}_1;\hat{\bm r}_2|\lambda_1\mu_1;\lambda_2\mu_2\rangle\\
=
\frac{C(\Lambda)}{R^2}\int d\hat{\bm r}_1Y^*_{l_1m_1}(\hat{\bm r}_1)Y^*_{l_2m_2}(\hat{\bm r}_1)Y^{}_{\lambda_1\mu_1}(\hat{\bm r}_1)Y^{}_{\lambda_2\mu_2}(\hat{\bm r}_1)\ .
\end{multline}
Usage of straightforward spherical harmonics algebra leads to the following expression:
{
\begin{multline}\label{eqn:4Ys}
\int d\hat{\bm r}\ Y^*_{l_1m_1}(\hat{\bm r})Y^*_{l_2m_2}(\hat{\bm r})Y^{}_{\lambda_1\mu_1}(\hat{\bm r})Y^{}_{\lambda_2\mu_2}(\hat{\bm r})=\frac{\sqrt{(2l_1+1)(2l_2+1))(2\lambda_1+1)(2\lambda_2+1)}}{4\pi}\\
\times\sum_{LM}(2L+1)
\begin{pmatrix}
l_1&l_2&L\\
-m_1&-m_2&M
\end{pmatrix}
\begin{pmatrix}
L&\lambda_1&\lambda_2\\
-M&\mu_1&\mu_2
\end{pmatrix}
\begin{pmatrix}
l_1&l_2&L\\
0&0&0
\end{pmatrix}
\begin{pmatrix}
L&\lambda_1&\lambda_2\\
0&0&0
\end{pmatrix}\\
\equiv \mathcal{Y}_4(l_1,m_1,l_2,m_2,\lambda_1,\mu_1,\lambda_2,\mu_2)\ .
\end{multline}
}  The triangle inequalities of the $3j-$symbols provide the following constraints: 
\begin{align*}
m_1+m_2=&M=\mu_1+\mu_2\\
\max(|l_1-l_2|,|\lambda_1-\lambda_2|)\le& L \le \min(l_1+l_2,\lambda_1+\lambda_2)\ .
\end{align*}
Note that the sum over $L,M$ in the $3j$-symbols in Eq.~\eqref{eqn:4Ys} does not allow a factorisation of terms between $l_1,m_1,l_2,m_2$ and $\lambda_1,\mu_1,\lambda_2,\mu_2$. To condense our expression a little we will use the abbreviation $\hat x\equiv 2x+1$ to end up with
{%\scriptsize
\begin{multline}\label{eqn:tiny}
\mathcal{Y}_{LM}(l_1,l_2,l'_1,l'_2)=\langle (l_1l_2)LM|\mathcal{Y}_4|(l_1'l_2')LM\rangle=\\
\sum_{\begin{matrix}m_1,m_2\\m'_1,m'_2\end{matrix}}\langle l_1m_1;l_2m_2|LM\rangle\mathcal{Y}_4(l_1,m_1,l_2,m_2,l'_1,m'_1,l'_2,m'_2)\langle l'_1m'_1;l'_2m'_2|LM\rangle
\\
=\frac{1}{4\pi}\sqrt{\hat{l_1}\hat{l_2}\hat{l'_1}\hat{l'_2}}\sum_{\mathcal{L}\mathcal{M}}\hat{\mathcal{L}}
\begin{pmatrix}
l_1 & l_2 & \mathcal{L}\\ 0 & 0 & 0
\end{pmatrix}
\begin{pmatrix}
l_1' & l_2' & \mathcal{L}\\ 0 & 0 & 0
\end{pmatrix}\\
\times
\left[\sum_{m_1m_2} \langle l_1m_1;l_2m_2|LM\rangle
\begin{pmatrix}
l_1 & l_2 & \mathcal{L}\\ -m_1 & -m_2 & \mathcal{M}
\end{pmatrix} \right]
\left[\sum_{m'_1m'_2} \langle l'_1m'_1;l'_2m'_2|LM\rangle
\begin{pmatrix}
l'_1 & l'_2 & \mathcal{L}\\ m'_1 & m'_2 & -\mathcal{M}
\end{pmatrix} \right]
\end{multline}} 
\noindent
We substituted $\mathcal{Y}_4(l_1,m_1,l_2,m_2,l'_1,m'_1,l'_2,m'_2)$ with Eq.~\eqref{eqn:4Ys} (and used the invariance of the 3$j$-symbols under cyclic permutation of indices). We further express the 3$j$-symbols using Clebsch-Gordan coefficients,
\begin{align}
\begin{pmatrix}
l_1 & l_2 & \mathcal{L}\\ -m_1 & -m_2 & \mathcal{M}
\end{pmatrix} &= \frac{(-1)^{l_1-l_2-\mathcal{M}}}{\sqrt{\hat{\mathcal{L}}}}\langle l_1,-m_1;l_2,-m_2|\mathcal{L},\mathcal{M}\rangle\\
\begin{pmatrix}
l'_1 & l'_2 & \mathcal{L}\\ m'_1 & m'_2 & -\mathcal{M}
\end{pmatrix} &=\frac{(-1)^{l_1'-l_2'+\mathcal{M}}}{\sqrt{\hat{\mathcal{L}}}}\langle l_1',m_1';l_2',m_2'|\mathcal{L},-\mathcal{M}\rangle.
\end{align}
Summing over the magnetic quantum numbers $m_1'$ and $m_2$ and applying the orthogonality of Clebsch-Gordan coefficients, the sum collapses to two Kronecker Deltas,
\begin{equation}
\sum_{m'_1m'_2} \langle l'_1m'_1;l'_2m'_2|LM\rangle\langle l_1',m_1';l_2',m_2'|\mathcal{L},-\mathcal{M}\rangle=\delta_{L,\mathcal{L}}\delta_{M,-\mathcal{M}}\ .
\end{equation}
Equation~Eq.~\eqref{eqn:tiny} therefore becomes
\begin{multline}\label{eqn:less tiny}
\mathcal{Y}_{LM}(l_1,l_2,l'_1,l'_2)=\langle (l_1l_2)LM|\mathcal{Y}_4|(l_1'l_2')LM\rangle=\\
\sum_{\begin{matrix}m_1,m_2\\m'_1,m'_2\end{matrix}}\langle l_1m_1;l_2m_2|LM\rangle\mathcal{Y}_4(l_1,m_1,l_2,m_2,l'_1,m'_1,l'_2,m'_2)\langle l'_1m'_1;l'_2m'_2|LM\rangle
\\
=(-1)^{l_1-l_2+l_1'-l_2'}\frac{1}{4\pi}\sqrt{\hat{l_1}\hat{l_2}\hat{l'_1}\hat{l'_2}}
\begin{pmatrix}
l_1 & l_2 &L\\ 0 & 0 & 0
\end{pmatrix}
\begin{pmatrix}
l_1' & l_2' & L\\ 0 & 0 & 0
\end{pmatrix}\\
\times
\left[\sum_{m_1m_2} \langle l_1m_1;l_2m_2|LM\rangle
\langle l_1,-m_1;l_2,-m_2|L,-M\rangle\right]
\end{multline}
Now we use the property $\langle l_1,-m_1;l_2,-m_2|L,-M\rangle=(-1)^{l_1-l_2-L}\langle l_1,m_1;l_2,m_2|L,M\rangle$, and the remaining sums over $m_1$ and $m_2$ give unity, leading to the final expression
\begin{align}\label{eqn:not so tiny anymore}
\mathcal{Y}_{LM}(l_1,l_2,l'_1,l'_2)
&=(-1)^{l_1'-l_2'-L}\frac{1}{4\pi}\sqrt{\hat{l_1}\hat{l_2}\hat{l'_1}\hat{l'_2}}
\begin{pmatrix}
l_1 & l_2 &L\\ 0 & 0 & 0
\end{pmatrix}
\begin{pmatrix}
l_1' & l_2' & L\\ 0 & 0 & 0
\end{pmatrix}\\
&=\frac{1}{4\pi}\sqrt{\hat{l_1}\hat{l_2}\hat{l'_1}\hat{l'_2}}
\begin{pmatrix}
l_1 & l_2 &L\\ 0 & 0 & 0
\end{pmatrix}
\begin{pmatrix}
l_1' & l_2' & L\\ 0 & 0 & 0
\end{pmatrix}\ ,
\end{align}
where we used the fact that for non-vanishing 3$j$ coefficient,  $l_1'+l_2'+L=$ must be even, which implies that the factor $(-1)^{l_1'-l_2'-L}=1$.

Inserting this result in Eq.~\eqref{eqn:ME} leaves us with the final form of the matrix element,
\begin{multline}\label{eqn:MELL}
\langle (l_1l_2)LM|\hat V_{12}
|(l'_1l'_2)LM\rangle
=\frac{C(\Lambda)}{R^2}\times \\
\sum_{\begin{matrix}m_1,m_2\\m'_1,m'_2\end{matrix}}\langle l_1m_1;l_2m_2|LM\rangle\mathcal{Y}_4(l_1,m_1,l_2,m_2,l'_1,m'_1,l'_2,m'_2)\langle l'_1m'_1;l'_2m'_2|LM\rangle\\
\equiv \frac{C(\Lambda)}{R^2}\mathcal{Y}_{LM}(l_1,l_2,l'_1,l'_2)\ .
\end{multline}
Note that the sums are restricted such that $m_1+m_2=M=m'_1+m'_2$.  
%So the analog of Eq.~\eqref{eqn:matrix equation} is
%\begin{equation}\label{eqn:matrix equation 2}
%\left(\mathbb{1}-\frac{4\pi}{\log(a\Lambda)}\mathbb{Y}_{LM}(x)\right)\cdot \bm\psi=0\ ,
%\end{equation}
%where the components of the matrix $\mathbb{Y}_{LM}(x)$ in this basis are
%\begin{equation}
%\mathbb{Y}_{LM}(x)^{l_1l_2}_{l'_1l'_2}=\frac{\mathcal{Y}_{LM}(l_1,l_2,l'_1,l'_2)}{l_1(l_1+1)+l_2%(l_2+1)-x}
% .
% \end{equation}
The matrix element $\mathcal{Y}_{LM}(l_1,l_2,l'_1,l'_2)$ can be analytically determined and is separable,
 \begin{multline}\label{eqn:YLM matrix element}
 \mathcal{Y}_{LM}(l_1,l_2,l'_1,l'_2)=\frac{1}{4\pi}\left[\sqrt{(2l_1+1)(2l_2+1)}
 \begin{pmatrix}
 l_1 & l_2 & L\\0 & 0 & 0
 \end{pmatrix}
 \right]\\
 \left[
 \sqrt{(2l_1'+1)(2l_2'+1)}
 \begin{pmatrix}
 l_1' & l_2' & L\\0 & 0 & 0
 \end{pmatrix}\right]
\end{multline}
This last relation, combined with~Eq.~\eqref{eqn:MELL}, gives the stated result in~Eq.~\eqref{eq:matrix_element}.

\nocite{*}
\bibliography{references}% Produces the bibliography via BibTeX.

\end{document}